\documentclass[final,3p,times,authoryear]{elsarticle}

\usepackage{amsmath}
\usepackage{amssymb}
\usepackage{amsthm}
\usepackage{booktabs}
\usepackage{graphicx}
\usepackage{microtype}
\usepackage{algorithm}
\usepackage{algorithmic}
\usepackage{tikz}
\usetikzlibrary{arrows.meta,positioning,calc}
\usepackage{pgfplots}
\pgfplotsset{compat=1.18}
\usepackage{hyperref}

\hypersetup{
  colorlinks=true,
  linkcolor=blue,
  citecolor=blue,
  urlcolor=blue
}

\makeatletter
\def\ps@pprintTitle{%
  \let\@oddhead\@empty
  \let\@evenhead\@empty
  \let\@oddfoot\@empty
  \let\@evenfoot\@empty
}
\makeatother

\newtheorem{theorem}{Theorem}
\newtheorem{lemma}{Lemma}
\newtheorem{corollary}{Corollary}
\theoremstyle{definition}
\newtheorem{definition}{Definition}
\newtheorem{remark}{Remark}

\begin{document}

\begin{frontmatter}

  \title{Exact Online $Q_n$ Estimation:\\
  Exploiting Sliding-Window Rank Invariance}

  \author{Robin Nunkesser}
  \ead{robin.nunkesser@hshl.de}
  \address{Department of Computer Science, Hamm-Lippstadt University
  of Applied Sciences, Marker Allee 76--78, 59063 Hamm, Germany}

  \begin{abstract}
    The $Q_n$ scale estimator \citep{RousseeuwCroux1993} combines a
    50\% finite-sample breakdown point with high Gaussian efficiency,
    but its computation requires selecting an order statistic among
    the $\binom{n}{2}$ pairwise distances. We study exact online
    $Q_n$ computation in a sliding window by exploiting temporal rank
    coherence between consecutive windows. Formalizing the multiset
    decomposition of surviving pairwise distances in terms of strict
    and weak rank counts, we derive a tie-exact dual-rank update that
    verifies whether the previous unscaled estimate remains valid in
    $\mathcal{O}(\log n)$ comparisons while maintaining the sorted
    window in $\mathcal{O}(n)$ time. When the target rank shifts by
    $m$ positions, a localized row-heap procedure initialized via a
    monotonic two-pointer frontier scan extracts the new exact order
    statistic in $\mathcal{O}(n + m\log n)$ comparisons without
    materializing the difference matrix, complemented by a hybrid
    fallback that executes frontier-guided implicit matrix selection
    in $\mathcal{O}(n\log n)$ time for larger rank shifts. On
    deterministic multi-regime stress streams, our implementation
    matches the offline
    batch reference.
    In repeated same-platform benchmarks on a synthetic Gaussian
    stream, our implementation achieves
    median internal throughputs from 2,220,000 updates/s ($n=25$) and
    1,140,000 updates/s ($n=51$) to 127,000 updates/s ($n=201$),
    outperforming both the exact \textsf{FQN} baseline ($2.5\times$
    to $6.4\times$) and the approximate \textsf{AFQN} sketch
    ($1.3\times$ to $2.7\times$) across all tested window sizes.
  \end{abstract}

  \begin{keyword}
    Robust statistics \sep $Q_n$ scale estimator \sep Data streams
    \sep Sliding window \sep Online algorithms \sep Algorithm engineering
  \end{keyword}

\end{frontmatter}

\section{Introduction}
\label{sec:intro}

The reliable, online estimation of variability in high-frequency time
series is a critical prerequisite in intensive care patient
monitoring \citep{NunkesserFriedGather2009}, algorithmic trading,
sensor telemetry, and industrial quality control \citep{ISO13528}. In
such streaming environments, observations arrive sequentially and are
routinely contaminated by measurement artifacts, transient sensor
spikes, or abrupt structural shifts. Classical moment-based
dispersion estimators such as the rolling sample variance or standard
deviation suffer from a finite-sample breakdown point of $0\%$: a
single unbounded outlier can inflate the scale estimate arbitrarily,
masking subsequent anomalies.

To overcome this vulnerability, \citet{RousseeuwCroux1993} introduced
the $Q_n$ scale estimator, defined for a univariate sample $X = (x_1,
\dots, x_n) \in \mathbb{R}^n$ as:
\begin{equation}
  \label{eq:qn_def}
  Q_n(X) = d \times d_n \times \{\!\{ |x_i - x_j| : 1 \le i < j \le n
  \}\!\}_{(k)},
\end{equation}
where $h = \lfloor n/2 \rfloor + 1$, $k = \binom{h}{2}$, and
$\{\!\{\cdot\}\!\}_{(k)}$ denotes the $k$-th order statistic of the
multiset of $N = \binom{n}{2}$ pairwise absolute differences. The
asymptotic consistency factor $d = 1 / (\sqrt{2}\,\Phi^{-1}(5/8))
\approx 2.219144466$ calibrates the estimator at Gaussian
distributions, while $d_n$ denotes a finite-sample bias correction
factor \citep{Fahmy2023}. Unlike the Median Absolute Deviation (MAD),
which achieves only $36.7\%$ Gaussian efficiency and measures
dispersion symmetrically around the sample median, $Q_n$ attains an
asymptotic Gaussian efficiency of $82.3\%$ while preserving the
maximal $50\%$ finite-sample breakdown point and requiring no prior
location estimate.

Throughout, algorithmic exactness refers to selecting the unscaled
$k$-th pairwise distance in the real-arithmetic model. The scale
factors are applied afterward as numerical constants. The
accompanying C core uses rounded tabulated values of $d_n$ for $n \le
12$ and a polynomial approximation for $n > 12$; its scaled output is
therefore not claimed to be an exact real-arithmetic value.

\subsection{Historical Trajectory and the Streaming Dilemma}

Evaluating Eq.~\eqref{eq:qn_def} naively requires $\mathcal{O}(n^2)$
space and $\mathcal{O}(n^2 \log n)$ time to construct and sort all
pairwise differences. In the offline batch setting,
\citet{CrouxRousseeuw1992} reduced this complexity by adapting the
implicit matrix selection algorithm of \citet{JohnsonMizoguchi1978},
computing $Q_n$ exactly in $\mathcal{O}(n \log n)$ time and
$\mathcal{O}(n)$ auxiliary memory.

In a streaming sliding-window scenario, where at time step $t+1$ the
oldest observation $x_{t-n+1}$ leaves the window of width $n$ and a
new observation $x_{t+1}$ enters, invoking a full $\mathcal{O}(n \log
n)$ batch solver from scratch at every step discards all structural
information from step $t$. Addressing this inefficiency has motivated
two decades of algorithmic research:
\begin{enumerate}
  \item \textbf{Dynamic Solution-Buffer Method:}
    \citet{SchettlingerNunkesserFried2008,NunkesserFriedGather2009}
    introduced the first exact
    online sliding-window algorithm for $Q_n$, extending earlier
    sliding-window paradigms for repeated median filters
    \citep{BernholtFriedGatherWegener2006}. By maintaining a dynamic
    solution buffer $\mathcal{B}$ of candidate differences around the
    $k$-th order statistic, they achieved an expected amortized update time of
    $\mathcal{O}(\log n)$ for a constant signal with stationary
    noise. However, whenever the true order statistic drifted outside
    the preallocated buffer (a buffer miss), the structure triggered
    a full $\mathcal{O}(n \log n)$ batch recomputation.
  \item \textbf{Worst-Case Linear Selection (\textsf{FQN}):}
    \citet{CafaroMellePulimenoEpicoco2021} proposed \textsf{FQN}
    (\emph{Fast Online $Q_n$}), achieving a deterministic worst-case
    update time of $\mathcal{O}(n)$ by maintaining the sorted window
    and applying the biselective matrix selection algorithm of
    \citet{MirzaianArjomandi1985} from scratch at every time step,
    without reusing the previous scale estimate $\theta^{(t)}$.
  \item \textbf{Approximate Streaming Estimation (\textsf{AFQN}):}
    \citet{EpicocoMelleCafaroPulimeno2022} introduced \textsf{AFQN}
    (\emph{Approximate Fast $Q_n$}), using a dynamically collapsing
    logarithmic quantile sketch to update the
    $n-1$ distances associated with each incoming and outgoing
    observation. On most tested synthetic distributions, the paper
    reports up to three times as many updates per second as
    \textsf{FQN}; the results depend on sketch configuration, and
    \textsf{FQN} is faster on the tested Poisson and Zipf
    distributions. As a sketch-based method, \textsf{AFQN} is approximate, and
    its accuracy depends on the sketch configuration.
  \item \textbf{Batch Constants Correction and Acceleration:} In
    Algorithm~1034 (2023), \citet{Fahmy2023} corrected an error in
    the asymptotic constant $d$ and recomputed the finite-sample
    factors $d_n$; Fahmy reports that the errors were also present in
    ISO~13528:2015 \citep{ISO13528}. On random binomial and normal
    samples of sizes $10$ to $100{,}000$, the updated batch algorithm
    achieved $1.3\times$ to $4.5\times$ the throughput of the
    \texttt{robustbase} \texttt{Qn} function. Fahmy also notes that
    the online framework of \citet{NunkesserFriedGather2009} could
    benefit from these algorithmic improvements.
\end{enumerate}

\subsection{Our Contribution: Exploiting Temporal Coherence}

This paper investigates how exploiting temporal rank coherence
between consecutive sliding windows can accelerate exact online $Q_n$
computation without maintaining explicit $\Theta(n^2)$ difference
buffers or sacrificing exactness to quantile sketches. Between $t$
and $t+1$, the sorted window changes by a single element replacement,
while the previous unscaled estimate $\theta^{(t)}$ either remains
the exact $k$-th order statistic or lies within a bounded rank neighborhood.

We combine tie-exact dual-rank invariance checks with localized
row-heap extraction and a contiguous-array implementation, and
evaluate our ANSI C implementation alongside native \textsf{FQN} and
\textsf{AFQN} reference codes on the same hardware. Our contributions are:
\begin{enumerate}
  \item \textbf{Tie-Exact $\mathcal{O}(\log n)$ Dual-Rank Update
      (Lemma~\ref{lem:invariance}, Theorem~\ref{thm:dual_rank} \&
    Corollary~\ref{cor:zero_search}):} Formalizing the multiset
    decomposition of the $\binom{n-1}{2}$ surviving pairwise
    distances in terms of strict and weak rank counting functions
    $(R_{<}, R_{\le})$, we bound the per-step rank perturbation by
    $|\Delta R| \le n-1$ and derive a closed-form update that
    verifies the previous scale estimate $\theta^{(t)}$ in the
    updated window using a constant number of binary searches
    ($\mathcal{O}(\log n)$ comparisons), identifying zero-search
    completions whenever the updated rank interval still brackets $k$.
  \item \textbf{Exact Localized Multi-Way Heap Adjustment
    (Theorem~\ref{thm:heap_correctness}):} When the target rank
    shifts by $m$ positions up to a configurable threshold $M_{\max}$
    ($1 \le m \le M_{\max}$), we prove that
    locating all row frontiers via a monotonic two-pointer scan and
    building a bounded min- or max-heap extracts the new exact $k$-th
    order statistic and updates its tie-exact ranks in $\mathcal{O}(n
    + m \log n)$ comparisons, completely avoiding full matrix search.
  \item \textbf{Hybrid Frontier-Guided Matrix Selection and Empirical
    Evaluation:} When $m > M_{\max}$, a hybrid fallback
    narrows each row's search interval to width at most $m \le n - 1$
    around the $\theta^{(t)}$ frontier and applies implicit
    weighted-median matrix selection
    \citep{JohnsonMizoguchi1978,CrouxRousseeuw1992} in $\mathcal{O}(n
    \log n)$ time and $\mathcal{O}(n)$ stack memory. We verify zero
    observed double-precision discrepancy against an independent
    offline batch reference on multi-regime stress streams and
    benchmark compiled C throughput for $n \in [25, 201]$. On a
    synthetic Gaussian stream, our implementation outperforms both
    native \textsf{FQN} ($2.5\times$ to $6.4\times$) and the
    approximate \textsf{AFQN} ($1.3\times$ to $2.7\times$) across
    every tested window size.
\end{enumerate}

The remainder of this paper is structured as follows:
Section~\ref{sec:theory} establishes the theoretical foundations and
proves the Dual-Rank Checking Theorem. Section~\ref{sec:algo}
presents the streaming algorithm, proves the correctness of localized
heap extraction, and analyzes its complexity and data layout.
Section~\ref{sec:verification} reports batch-agreement checks under
synthetic stress regimes and throughput measurements.
Section~\ref{sec:software} describes our C/R/Python implementations,
and Section~\ref{sec:conclusion} concludes.

\section{Theoretical Foundations and Dual-Rank Verification}
\label{sec:theory}

Let $W_t = (x_{t-n+1}, \dots, x_t) \in \mathbb{R}^n$ denote the
active sliding window of width $n \ge 2$ at time step $t \ge n$, and
let $x_{(1)} \le x_{(2)} \le \dots \le x_{(n)}$ denote its sorted
order statistics. The implicit upper-triangular difference matrix
generates the multiset of $N = \binom{n}{2}$ non-negative pairwise distances:
\begin{equation}
  \label{eq:multiset_d}
  D(W_t) = \{\!\{ x_{(j)} - x_{(i)} \mid 1 \le i < j \le n \}\!\},
  \quad |D(W_t)| = N = \frac{n(n-1)}{2}.
\end{equation}
The doubled braces denote a multiset: each pair of observation
occurrences contributes one distance, including when several pairs
have the same value. The symbol $\uplus$ denotes multiset union,
which adds multiplicities.
For a multiset $M$, let $C_{<}(M,\theta)$ and $C_{\le}(M,\theta)$
count the occurrences below or at most $\theta$, respectively, with
multiplicity.
Let $h = \lfloor n/2 \rfloor + 1$ and let $k = \binom{h}{2} =
h(h-1)/2 \in \{1, \dots, N\}$ denote the 1-based target rank of the
unscaled $Q_n$ statistic.

Because empirical data streams (such as quantized sensor readings or
discrete financial ticks) frequently contain duplicate values or tied
differences, a single rank count is insufficient to uniquely
characterize order statistics. For any candidate threshold $\theta
\ge 0$, we therefore define the \emph{strict rank counting function}
$R_{<}(\theta; W_t)$ and the \emph{weak rank counting function}
$R_{\le}(\theta; W_t)$:
\begin{align}
  R_{<}(\theta; W_t) &= \left| \left\{ (i, j) \;\middle|\; 1 \le i <
  j \le n, \; x_{(j)} - x_{(i)} < \theta \right\} \right|,
  \label{eq:def_r_lt} \\
  R_{\le}(\theta; W_t) &= \left| \left\{ (i, j) \;\middle|\; 1 \le i
  < j \le n, \; x_{(j)} - x_{(i)} \le \theta \right\} \right|.
  \label{eq:def_r_le}
\end{align}

\begin{definition}[Tie-Exact $k$-th Order Statistic]
  \label{def:order_stat}
  A non-negative scalar $\theta^* \ge 0$ is the exact $k$-th order
  statistic of $D(W_t)$, denoted $\theta^* = D(W_t)_{(k)}$, if and
  only if the dual-rank bracketing condition holds:
  \begin{equation}
    \label{eq:order_stat_condition}
    R_{<}(\theta^*; W_t) < k \quad \text{and} \quad R_{\le}(\theta^*;
    W_t) \ge k.
  \end{equation}
\end{definition}

Note that Definition~\ref{def:order_stat} handles arbitrary
multiplicity of ties seamlessly: if $\theta^*$ appears $c \ge 1$
times in $D(W_t)$, then $R_{\le}(\theta^*; W_t) - R_{<}(\theta^*;
W_t) = c$, and $\theta^*$ occupies all 1-based ranks in the closed
integer interval $[R_{<}(\theta^*; W_t) + 1, \, R_{\le}(\theta^*; W_t)]$.

\subsection{Sliding-Window Distance Invariance and Bounded Rank Drift}

When advancing from time $t$ to $t+1$, the oldest observation
occurrence $x_{\text{old}} = x_{t-n+1}$ is evicted and the newest
observation $x_{\text{new}} = x_{t+1}$ is ingested. Let
$W_{\text{shared}}$ be the multiset of the $n-1$ surviving
observation occurrences, so that $W_t = W_{\text{shared}} \uplus
\{x_{\text{old}}\}$ and $W_{t+1} = W_{\text{shared}} \uplus
\{x_{\text{new}}\}$. As exploited by earlier online buffer and sketch
updates
\citep{SchettlingerNunkesserFried2008,NunkesserFriedGather2009,EpicocoMelleCafaroPulimeno2022},
only the $n-1$ distances involving $x_{\text{old}}$ leave the pair
multiset and $n-1$ distances involving $x_{\text{new}}$ enter. The
following lemma formalizes how this decomposition bounds the per-step
shift of the dual-rank counting functions $R_{<}$ and $R_{\le}$
without storing the pairwise differences explicitly.

\begin{lemma}[Sliding-Window Multiset Decomposition and Rank Drift Bound]
  \label{lem:invariance}
  Let $W_{t+1} = W_{\text{shared}} \uplus \{x_{\text{new}}\}$ be
  obtained from $W_t = W_{\text{shared}} \uplus \{x_{\text{old}}\}$
  by a single sliding-window transition. Then:
  \begin{enumerate}
    \item Let $D_{\text{evict}}$ and $D_{\text{ingest}}$ be the
      multisets of distances from the evicted and incoming
      observations, respectively, to each occurrence in
      $W_{\text{shared}}$. Then:
      \begin{align*}
        D(W_t) &= D(W_{\text{shared}}) \uplus D_{\text{evict}}, \\
        D(W_{t+1}) &= D(W_{\text{shared}}) \uplus D_{\text{ingest}}.
      \end{align*}
      Both change multisets have cardinality $n-1$ counting
      multiplicities, while the invariant core $D(W_{\text{shared}})$ contains:
      \begin{equation}
        |D(W_{\text{shared}})| = \binom{n-1}{2} = \left(1 -
        \frac{2}{n}\right) \binom{n}{2}
      \end{equation}
      unchanged pairwise distances.
    \item For any fixed threshold $\theta \ge 0$, the strict and weak
      rank shifts between $W_t$ and $W_{t+1}$ satisfy:
      \begin{equation}
        \label{eq:drift_bound}
        -(n - 1) \le R_{<}(\theta; W_{t+1}) - R_{<}(\theta; W_t) \le
        n - 1, \quad -(n - 1) \le R_{\le}(\theta; W_{t+1}) -
        R_{\le}(\theta; W_t) \le n - 1.
      \end{equation}
  \end{enumerate}
\end{lemma}

\noindent\textit{Proof.} See Appendix~\ref{app:proofs}.

Lemma~\ref{lem:invariance} reflects that the fraction of unchanged
pairwise-distance occurrences is high: for $n=51$, $\rho_{51} =
96.1\%$; for $n=101$, $\rho_{101} = 98.0\%$; and for $n=501$,
$\rho_{501} = 99.6\%$. Although $N = \binom{n}{2} = \Theta(n^2)$, the
rank of any fixed value $\theta$ can drift by at most
$\mathcal{O}(n)$ positions per step. This deterministic bound alone
does not imply a smaller expected drift for stochastic input; the
empirical drift profile for the tested stream is reported in
Section~\ref{sec:algo}.

\subsection{The \texorpdfstring{$\mathcal{O}(\log n)$}{O(log n)}
Dual-Rank Checking Theorem}

We now show that the exact ranks $R_{<}(\theta; W_{t+1})$ and
$R_{\le}(\theta; W_{t+1})$ of the previous unscaled estimate $\theta
= \theta^{(t)}$ can be updated using only $\mathcal{O}(\log n)$ comparisons.

\begin{theorem}[Logarithmic Dual-Rank Update]
  \label{thm:dual_rank}
  Let $\theta = \theta^{(t)}$ be the unscaled $Q_n$ order statistic
  at time $t$ with known rank counts $R_{<}(\theta; W_t)$ and
  $R_{\le}(\theta; W_t)$. Upon the sliding-window transition that
  evicts occurrence $x_{\text{old}}$ and inserts $x_{\text{new}}$,
  the exact rank counts of $\theta$ in $W_{t+1}$ satisfy:
  \begin{align}
    R_{<}(\theta; W_{t+1}) &= R_{<}(\theta; W_t) -
    \Delta_{<}(x_{\text{old}}, \theta; W_t) +
    \Delta_{<}(x_{\text{new}}, \theta; W_{t+1}), \label{eq:update_lt} \\
    R_{\le}(\theta; W_{t+1}) &= R_{\le}(\theta; W_t) -
    \Delta_{\le}(x_{\text{old}}, \theta; W_t) +
    \Delta_{\le}(x_{\text{new}}, \theta; W_{t+1}), \label{eq:update_le}
  \end{align}
  where for any window $W$ with sorted representation $x_{(1)} \le
  \dots \le x_{(n)}$ and distinguished observation occurrence $v$ at
  sorted index $p$, the marginal pair counts $\Delta_{<}(v, \theta;
  W) = |\{ j \ne p : |x_{(j)} - v| < \theta \}|$ and $\Delta_{\le}(v,
  \theta; W) = |\{ j \ne p : |x_{(j)} - v| \le \theta \}|$ are given
  in closed form by:
  \begin{align}
    \Delta_{<}(v, \theta; W) &= \max\!\left(0, \; U_{\ge}(v + \theta)
    - U_{>}(v - \theta) - \mathbf{1}_{\{\theta > 0\}}\right),
    \label{eq:bsearch_lt} \\
    \Delta_{\le}(v, \theta; W) &= \max\!\left(0, \; U_{>}(v + \theta)
    - U_{\ge}(v - \theta) - 1\right), \label{eq:bsearch_le}
  \end{align}
  where $U_{\ge}(z) = \min(\{ j \in \{1, \dots, n\} : x_{(j)} \ge z
  \} \cup \{n+1\})$ and $U_{>}(z) = \min(\{ j \in \{1, \dots, n\} :
  x_{(j)} > z \} \cup \{n+1\})$ denote standard lower- and
  upper-bound binary search indices, computable in $\mathcal{O}(\log
  n)$ comparisons.
\end{theorem}

\noindent\textit{Proof.} Because $x_{(1)} \le \dots \le x_{(n)}$ is
sorted, counting pairs within distance $< \theta$ or $\le \theta$ of
any value $v$ reduces to locating the contiguous index interval
around $v \pm \theta$ via binary search and excluding the self-pair;
the full proof is given in Appendix~\ref{app:proofs}.

\begin{corollary}[Zero-Search Rank-Invariant Update]
  \label{cor:zero_search}
  If the updated rank counts from Theorem~\ref{thm:dual_rank} satisfy
  $R_{<}(\theta^{(t)}; W_{t+1}) < k$ and $R_{\le}(\theta^{(t)};
  W_{t+1}) \ge k$, then by Definition~\ref{def:order_stat},
  $\theta^{(t+1)} = \theta^{(t)}$ is the exact $k$-th order statistic
  of $D(W_{t+1})$. Consequently, $Q_n(W_{t+1}) = Q_n(W_t)$ is
  verified in $\mathcal{O}(\log n)$ comparisons with zero matrix
  candidate evaluations.
\end{corollary}

\section{Algorithmic Architecture and Correctness Analysis}
\label{sec:algo}

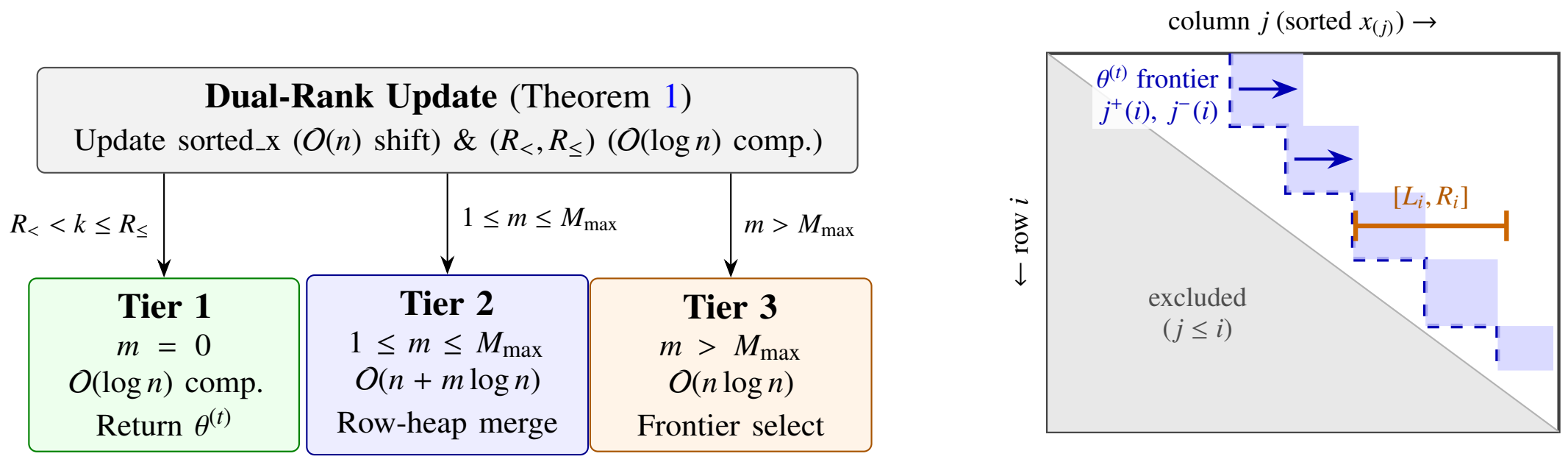
\begin{figure}[tbp]
  \centering
  \begin{tikzpicture}[
      >=Stealth,
      font=\small,
      box/.style={draw, rounded corners=2pt, align=center, inner
      sep=4pt, minimum height=0.85cm},
      tier1/.style={box, fill=green!7, draw=green!50!black},
      tier2/.style={box, fill=blue!7, draw=blue!55!black},
      tier3/.style={box, fill=orange!9, draw=orange!65!black},
      hdr/.style={box, fill=gray!10, draw=black!70, text width=7.1cm}
    ]
    \node[hdr] (init) at (0, 3.05) {\textbf{Dual-Rank Update}
      (Theorem~\ref{thm:dual_rank})\\
      \footnotesize Update $\text{sorted\_x}$ ($\mathcal{O}(n)$ shift)
    \& $(R_{<}, R_{\le})$ ($\mathcal{O}(\log n)$ comp.)};

    \node[tier1, text width=2.15cm] (t1) at (-2.55, 0.85) {\textbf{Tier 1}\\
      \footnotesize $m = 0$\\
      \footnotesize $\mathcal{O}(\log n)$ comp.\\
    \footnotesize Return $\theta^{(t)}$};

    \node[tier2, text width=2.25cm] (t2) at (0.0, 0.85) {\textbf{Tier 2}\\
      \footnotesize $1 \le m \le M_{\max}$\\
      \footnotesize $\mathcal{O}(n + m\log n)$\\
    \footnotesize Row-heap merge};

    \node[tier3, text width=2.25cm] (t3) at (2.55, 0.85) {\textbf{Tier 3}\\
      \footnotesize $m > M_{\max}$\\
      \footnotesize $\mathcal{O}(n\log n)$\\
    \footnotesize Frontier select};

    \draw[->] (init.south -| t1.north) -- node[left,
    font=\scriptsize, align=right] {$R_{<}<k\le R_{\le}$} (t1.north);
    \draw[->] (init.south) -- node[right, font=\scriptsize, pos=0.45]
    {$1 \le m \le M_{\max}$} (t2.north);
    \draw[->] (init.south -| t3.north) -- node[right,
    font=\scriptsize, align=left] {$m > M_{\max}$} (t3.north);

    \node[font=\footnotesize\bfseries] at (0, -0.35) {(a) Three-tier
    routing by rank drift $m$};

    \begin{scope}[shift={(5.4, 0.25)}]
      \draw[thick, draw=black!75] (0,0) rectangle (4.6, 3.4);
      \fill[gray!18] (0,3.4) -- (0,0) -- (4.6,0) -- cycle;
      \draw[gray!60] (0,3.4) -- (4.6,0);
      \node[gray!60!black, font=\scriptsize, align=center] at (1.35,
      1.05) {excluded\\($j \le i$)};

      \draw[very thick, dashed, blue!70!black]
      (1.65, 3.4) -- (1.65, 2.75) --
      (2.15, 2.75) -- (2.15, 2.15) --
      (2.75, 2.15) -- (2.75, 1.55) --
      (3.40, 1.55) -- (3.40, 0.95) --
      (4.05, 0.95) -- (4.05, 0.55);

      \fill[blue!18, opacity=0.8]
      (1.65, 3.4) -- (2.30, 3.4) -- (2.30, 2.75) --
      (2.80, 2.75) -- (2.80, 2.15) --
      (3.40, 2.15) -- (3.40, 1.55) --
      (4.05, 1.55) -- (4.05, 0.95) --
      (4.55, 0.95) -- (4.55, 0.55) --
      (4.05, 0.55) -- (4.05, 0.95) --
      (3.40, 0.95) -- (3.40, 1.55) --
      (2.75, 1.55) -- (2.75, 2.15) --
      (2.15, 2.15) -- (2.15, 2.75) --
      (1.65, 2.75) -- cycle;

      \draw[very thick, orange!80!black, |-|] (2.75, 1.85) -- (4.15, 1.85);
      \node[orange!70!black, font=\scriptsize, anchor=south] at
      (3.45, 1.88) {$[L_i, R_i]$};

      \draw[->, blue!70!black, thick] (1.72, 3.08) -- (2.25, 3.08);
      \draw[->, blue!70!black, thick] (2.22, 2.45) -- (2.75, 2.45);

      \node[font=\scriptsize, anchor=south] at (2.3, 3.42) {column
      $j$ (sorted $x_{(j)}$) $\rightarrow$};
      \node[font=\scriptsize, rotate=90, anchor=south] at (-0.05,
      1.7) {$\leftarrow$ row $i$};
      \node[blue!70!black, font=\scriptsize, align=right,
        anchor=east, fill=white, fill opacity=0.85, text opacity=1,
      inner sep=1pt] at (1.58, 3.03) {$\theta^{(t)}$
      frontier\\$j^+(i),\, j^-(i)$};

      \node[font=\footnotesize\bfseries] at (2.3, -0.60) {(b) Action
      on implicit matrix $D_{i,j}$};
    \end{scope}
  \end{tikzpicture}
  \caption{Overview of the three-tier online $Q_n$ update. (a)~After
    updating the sorted window and dual ranks $(R_{<}, R_{\le})$, steps
    with $m=0$ finish in Tier~1, small shifts ($1 \le m \le M_{\max}$)
    are resolved by a row-heap (Tier~2), and larger shifts ($m >
    M_{\max}$) use frontier-guided implicit matrix selection (Tier~3).
    (b)~Monotonicity of $D_{i,j} = x_{(j)} - x_{(i)}$ induces a
    non-decreasing $\theta^{(t)}$-frontier that seeds the Tier~2 heap
  band and restricts each row to $[L_i, R_i]$ of width at most $m$ in Tier~3.}
  \label{fig:three_tier_architecture}
\end{figure}

Figure~\ref{fig:three_tier_architecture} illustrates the three-tier
routing architecture and its geometric action on the implicit upper
triangular difference matrix $D_{i,j} = x_{(j)} - x_{(i)}$.
Algorithm~\ref{alg:online_qn} specifies the exact update once the
window is full and $\theta$ and its ranks have been initialized.
Startup filling and initialization are separate. The state maintains a
circular buffer $\text{ring}[0\dots n-1]$ for $\mathcal{O}(1)$ FIFO
eviction, its next-write index $\text{ring\_idx}$, a contiguous
sorted array $\text{sorted\_x}[0\dots n-1]$, the unscaled order
statistic $\theta$, and its exact dual ranks $(R_{<}, R_{\le})$.

\begin{algorithm}[t]
  \caption{Exact Online $Q_n$ Streaming Update}
  \label{alg:online_qn}
  \begin{algorithmic}[1]
    \REQUIRE Full-window state $(n, k, d, d_n, M_{\max}, \text{ring},
    \text{ring\_idx}, \text{sorted\_x}, \theta, R_{<}, R_{\le})$ with
    valid ranks, and incoming observation $x_{\text{new}}$
    \ENSURE Scaled estimate $Q_n^{(t+1)}$ based on the exact unscaled
    order statistic, and updated state
    \STATE $x_{\text{old}} \leftarrow \text{ring}[\text{ring\_idx}]$;
    $\text{ring}[\text{ring\_idx}] \leftarrow x_{\text{new}}$;
    $\text{ring\_idx} \leftarrow (\text{ring\_idx} + 1) \bmod n$
    \STATE $(\text{lost}_{<}, \text{lost}_{\le}) \leftarrow
    \textsc{CountPairs}(\text{sorted\_x}, n, x_{\text{old}}, \theta)$
    \COMMENT{Eqs.~\eqref{eq:bsearch_lt}--\eqref{eq:bsearch_le}:
    $\mathcal{O}(\log n)$ comparisons}
    \STATE Update $\text{sorted\_x}$: locate $x_{\text{old}}$ and
    $x_{\text{new}}$ via binary search ($\mathcal{O}(\log n)$) and
    shift via \texttt{memmove} ($\mathcal{O}(n)$)
    \STATE $(\text{gained}_{<}, \text{gained}_{\le}) \leftarrow
    \textsc{CountPairs}(\text{sorted\_x}, n, x_{\text{new}}, \theta)$
    \COMMENT{Eqs.~\eqref{eq:bsearch_lt}--\eqref{eq:bsearch_le}:
    $\mathcal{O}(\log n)$ comparisons}
    \STATE $R_{<} \leftarrow R_{<} - \text{lost}_{<} +
    \text{gained}_{<}$; \quad $R_{\le} \leftarrow R_{\le} -
    \text{lost}_{\le} + \text{gained}_{\le}$
    \IF{$R_{<} < k \text{ \textbf{and} } R_{\le} \ge k$}
    \STATE \textbf{return} $\theta \times d \times d_n$ \COMMENT{Tier
    1 (Zero-Step Hit): $\theta^{(t+1)} = \theta^{(t)}$}
    \ELSIF{$R_{\le} < k$}
    \STATE $m \leftarrow k - R_{\le}$
    \IF{$m \le M_{\max}$}
    \STATE $(\theta, R_{<}, R_{\le}) \leftarrow
    \textsc{ExtractNextMinCandidates}(\text{sorted\_x}, n, k, \theta,
    m)$ \COMMENT{Tier 2a (Min-Heap): $\mathcal{O}(n + m \log n)$}
    \ELSE
    \STATE $(\theta, R_{<}, R_{\le}) \leftarrow
    \textsc{FrontierGuidedMatrixSelect}(\text{sorted\_x}, n, k,
    \theta, R_{<}, R_{\le})$ \COMMENT{Tier 3 (Hybrid): $\mathcal{O}(n \log n)$}
    \ENDIF
    \ELSE
    \STATE $m \leftarrow R_{<} - k + 1$
    \IF{$m \le M_{\max}$}
    \STATE $(\theta, R_{<}, R_{\le}) \leftarrow
    \textsc{ExtractPrevMaxCandidates}(\text{sorted\_x}, n, k, \theta,
    m)$ \COMMENT{Tier 2b (Max-Heap): $\mathcal{O}(n + m \log n)$}
    \ELSE
    \STATE $(\theta, R_{<}, R_{\le}) \leftarrow
    \textsc{FrontierGuidedMatrixSelect}(\text{sorted\_x}, n, k,
    \theta, R_{<}, R_{\le})$ \COMMENT{Tier 3 (Hybrid): $\mathcal{O}(n \log n)$}
    \ENDIF
    \ENDIF
    \STATE \textbf{return} $\theta \times d \times d_n$
  \end{algorithmic}
\end{algorithm}

\subsection{Correctness of Localized Priority-Queue Rank Adjustment}

When Corollary~\ref{cor:zero_search} is not triggered, $\theta^{(t)}$
is no longer the $k$-th order statistic. To balance localized heap
extractions ($\mathcal{O}(n + m \log n)$) against implicit matrix
selection ($\mathcal{O}(n \log n)$), we introduce a configurable
shift threshold $M_{\max} \in \mathbb{N}$ (set to $M_{\max} = 32$ by
default) that governs the switch between Tier~2 and Tier~3. Empirical
profiling across 100,000 updates confirms that the rank drift $m$ is
tightly concentrated around zero: for $n = 51$, $92.6\%$ of all
updates require only a small local rank adjustment $1 \le m \le
M_{\max}$ (with median drift $m = 5$), while $7.4\%$ are instantaneous
zero-step hits ($m = 0$).

Rather than invoking a global matrix selection algorithm from
scratch, Tier~2 of Algorithm~\ref{alg:online_qn} resolves any drift
$m \le M_{\max}$ by performing a localized multi-way merge across the
rows of the implicit sorted difference matrix.

\begin{theorem}[Exactness and Linear-Frontier Complexity of Row-Heap Extraction]
  \label{thm:heap_correctness}
  Suppose that after updating $(R_{<}, R_{\le})$ at $\theta =
  \theta^{(t)}$ in $W_{t+1}$, either $R_{\le}(\theta; W_{t+1}) < k$
  (upward drift $m = k - R_{\le} \ge 1$) or $R_{<}(\theta; W_{t+1})
  \ge k$ (downward drift $m = R_{<} - k + 1 \ge 1$). Then:
  \begin{enumerate}
    \item \textbf{Upward Drift ($R_{\le} < k$):} For each row $i \in
      \{1, \dots, n-1\}$ for which the set is nonempty, let $j^+(i) =
      \min\{ j \in \{i+1, \dots, n\} : x_{(j)} - x_{(i)} > \theta
      \}$. Initializing a binary min-heap $\mathcal{H}_{\min}$ with
      the frontier triples $(x_{(j^+(i))} - x_{(i)}, \, i, \,
      j^+(i))$ and performing $m$ extract-min steps---where popping
      $(v, i, j)$ inserts its row successor $(x_{(j+1)} - x_{(i)}, \,
      i, \, j+1)$ if $j < n$---yields at the $m$-th extraction the
      exact $k$-th order statistic $\theta^{(t+1)} = D(W_{t+1})_{(k)}$.
    \item \textbf{Downward Drift ($R_{<} \ge k$):} Symmetrically, for
      each row $i \in \{1, \dots, n-1\}$ for which the set is
      nonempty, let $j^-(i) = \max\{ j \in \{i+1, \dots, n\} :
      x_{(j)} - x_{(i)} < \theta \}$. Initializing a binary max-heap
      $\mathcal{H}_{\max}$ with the frontier triples $(x_{(j^-(i))} -
      x_{(i)}, \, i, \, j^-(i))$ and performing $m = R_{<} - k + 1$
      extract-max steps---replacing $(v, i, j)$ with its row
      predecessor $(x_{(j-1)} - x_{(i)}, \, i, \, j-1)$ if $j - 1 >
      i$---yields at the $m$-th extraction $\theta^{(t+1)} = D(W_{t+1})_{(k)}$.
  \end{enumerate}
  Both procedures execute in $\mathcal{O}(n + m \log n)$ comparisons
  and $\mathcal{O}(n)$ stack memory.
\end{theorem}

\noindent\textit{Proof.} Monotonicity of $x_{(j)} - x_{(i)}$ in both
$i$ and $j$ ensures that the frontier indices $j^+(i)$ and $j^-(i)$
are non-decreasing in $i$ (located in $\mathcal{O}(n)$ via a single
two-pointer scan) and that a Floyd bottom-up heap performs an exact
$m$-step multi-way merge in $\mathcal{O}(n + m \log n)$ comparisons;
the full inductive proof is given in Appendix~\ref{app:proofs}.

Furthermore, Tier~2 updates $(R_{<}, R_{\le})$ at $\theta^{(t+1)}$ in
$\mathcal{O}(1)$ time whenever no tie extends past the $m$-th
extraction: if $c_{\text{before}} \ge 0$ of the first $m-1$ extracted
values equal $\theta^{(t+1)}$ and the next heap top after the $m$-th
extraction is strictly greater than $\theta^{(t+1)}$ (Case~1) or
strictly less than $\theta^{(t+1)}$ (Case~2), then $(R_{<}, R_{\le})
= (k - 1 - c_{\text{before}}, \, k)$ in Case~1 and $(k - 1, \, k +
c_{\text{before}})$ in Case~2. Only when a tie spans across the
$m$-th extraction does the implementation invoke a monotonic
two-pointer scan (\texttt{count\_all\_pairs}) across
\texttt{sorted\_x} in $\mathcal{O}(n)$ operations.

\subsection{Hybrid Tier-3 Matrix Selection and Contiguous Data Layout}

When the rank shift $m$ exceeds $M_{\max}$ (Tier~3),
Algorithm~\ref{alg:online_qn} invokes a hybrid
\textsc{FrontierGuidedMatrixSelect} procedure that combines our known
rank bounds at $\theta^{(t)}$ with the implicit matrix selection
paradigm of \citet{JohnsonMizoguchi1978} and \citet{CrouxRousseeuw1992}:
\begin{itemize}
  \item \textbf{Frontier-Guided Band Narrowing:} By
    Lemma~\ref{lem:invariance}, even when $m > M_{\max}$, the rank
    drift satisfies $m \le n - 1$. In Case~1 ($R_{\le}(\theta^{(t)})
    < k$), the $m$-th smallest element of $D_{>}(\theta^{(t)})$ in
    row $i$ must lie within the column interval $[L_i, R_i] =
    [j^+(i), \, \min(n, j^+(i) + m - 1)]$; symmetrically, in Case~2
    ($R_{<}(\theta^{(t)}) \ge k$), it must lie within $[L_i, R_i] =
    [\max(i+1, j^-(i) - m + 1), \, j^-(i)]$. Initializing $[L_i,
    R_i]$ via the $\mathcal{O}(n)$ two-pointer frontier scan
    immediately restricts each row to at most $m \le n - 1$
    candidates before running weighted-median bisection rounds in
    $\mathcal{O}(n \log n)$ time and $\mathcal{O}(n)$ stack memory.
  \item \textbf{Pair-Space Verification vs.\ Contiguous Array
    Maintenance:} Whereas batch selection
    \citep{CrouxRousseeuw1992,Fahmy2023} and \textsf{FQN}
    \citep{CafaroMellePulimenoEpicoco2021} execute full matrix
    selection on every step, our Dual-Rank Theorem reduces pair-space
    verification to $\mathcal{O}(\log n)$ binary search comparisons
    on Tier~1 hits and $\mathcal{O}(n + m \log n)$ on Tier~2
    adjustments. The implementation stores \texttt{sorted\_x} in a
    contiguous array updated via binary search and \texttt{memmove}
    ($\mathcal{O}(n)$ shifts) and evaluates all frontier and
    marginal-count predicates directly on differences $x_{(j)} - x_{(i)}$.
\end{itemize}

\begin{remark}[Direct Analogue for the Hodges--Lehmann Location Estimator]
  \label{rem:hodges_lehmann}
  The entire three-tier architecture transfers directly to the
  sliding-window Hodges--Lehmann location estimator, $\text{med}_{1 \le
  i < j \le n} (x_i + x_j)/2$ (or including $i = j$). Replacing the
  pair differences $x_{(j)} - x_{(i)}$ with the sums $x_{(i)} +
  x_{(j)}$ preserves Lemma~\ref{lem:invariance}; in
  Theorem~\ref{thm:dual_rank}, counting pairs with $x_{(j)} + v <
  2\theta$ or $\le 2\theta$ reduces to single-sided binary searches
  at $2\theta - v$ in $\mathcal{O}(\log n)$ comparisons. Because
  $x_{(i)} + x_{(j)}$ is monotonically non-decreasing in both $i$ and
  $j$, the frontier indices $j^+(i)$ and $j^-(i)$ become monotonically
  non-increasing in $i$, so the $\mathcal{O}(n)$ two-pointer scan
  simply traverses $j$ in the opposite direction while Tier~2
  ($\mathcal{O}(n + m\log n)$) and Tier~3 ($\mathcal{O}(n\log n)$)
  remain unchanged.
\end{remark}

\section{Batch Agreement and Throughput Benchmarks}
\label{sec:verification}

Both the single-run execution-tier benchmark and the repeated
head-to-head benchmark were executed on an Apple M3 Pro running
macOS~26.6.2 (arm64) using Apple Clang~21.0.0 with \texttt{-O3}.

\subsection{Batch-Agreement Check Under Multi-Regime Stress}

We compared the online implementation with an independent naive
$\mathcal{O}(n^2\log n)$ offline batch reference on deterministic $T
= 2,000$-step synthetic streams for window sizes $n \in \{31, 51,
101\}$. Each stream traversed four regimes:
\begin{enumerate}
  \item \textbf{Stationary Gaussian Regime ($t \in [0, 299]$):}
    Standard normal observations $x_t \sim \mathcal{N}(0, 1)$.
  \item \textbf{Adversarial Outlier Burst ($t \in [300, 329]$):} A
    contiguous block of $30$ extreme outliers shifted by $+50\sigma$
    ($x_t \sim \mathcal{N}(50, 1)$), testing breakdown resilience and
    rapid rank transitions.
  \item \textbf{Abrupt Structural Level Shift ($t \in [600, 999]$):}
    An instantaneous step-change in mean ($x_t \sim \mathcal{N}(10,
    1)$), testing bimodal window transitions.
  \item \textbf{Heavy-Tailed Regime ($t \in [1200, 1499]$):}
    Cauchy-like extreme kurtosis drawn from a Student-$t_3$
    distribution ($x_t \sim t_3$).
\end{enumerate}

All unlisted steps, namely $t \in [330, 599] \cup [1000, 1199] \cup
[1500, 1999]$, also use the stationary Gaussian baseline $x_t \sim
\mathcal{N}(0,1)$. The streams are generated with the Box--Muller
transform from the C pseudorandom generator initialized once with
\texttt{srand(12345)}; its state is consumed sequentially across the
three window-size runs.

At every time step with a full window, the online output
$Q_n^{\text{online}}(W_t)$ was compared against a fresh batch
evaluation $Q_n^{\text{batch}}(W_t)$ computed from Eq.~\eqref{eq:qn_def}.

\begin{table}[htbp]
  \centering
  \caption{Numerical agreement of Algorithm~\ref{alg:online_qn} with
    the offline batch reference across $T = 2,000$ multi-regime steps
    (Gaussian, $+50\sigma$ outlier burst, $+10\sigma$ level shift, and
  Student-$t_3$ heavy tails).}
  \label{tab:verification}
  \small
  \setlength{\tabcolsep}{4.5pt}
  \begin{tabular}{rrrrrrr}
    \toprule
    \textbf{Window $n$} & \textbf{Pairs $\binom{n}{2}$} &
    \textbf{Updates} & \textbf{Max Error $|Q_n^{\text{online}} -
    Q_n^{\text{batch}}|$} & \textbf{Zero-Step Hits} & \textbf{Heap
    Hits} & \textbf{Fallback Hits} \\
    \midrule
    31  &   465 & 1,969 & 0 (0 mismatches) & 10.26\% & 89.74\% & 0.00\% \\
    51  & 1,275 & 1,949 & 0 (0 mismatches) &  8.21\% & 91.79\% & 0.00\% \\
    101 & 5,050 & 1,899 & 0 (0 mismatches) &  4.21\% & 91.89\% & 3.90\% \\
    \bottomrule
  \end{tabular}
\end{table}

As documented in Table~\ref{tab:verification}, the online and batch
implementations produced identical double-precision outputs on all
$5,817$ reported sliding-window updates, as well as on exhaustive
small-alphabet, tie-heavy, and rounding-boundary unit tests.

\subsection{Single-Run Throughput and Execution-Tier Profile}

To measure throughput and execution-tier distributions, the harness
generated $T = 100,000$ Gaussian observations for each window size $n
\in \{25, 35, 51, 75, 101, 151, 201\}$ with heap drift threshold
$M_{\max} = 32$. It used the first $n$ observations to initialize the
window and timed the remaining $T-n$ updates with the C \texttt{clock()} timer.

\begin{table}[htbp]
  \centering
  \caption{Streaming throughput, per-update latency, and execution
    tier breakdown in compiled ANSI C ($T=100,000$ observations; $T-n$
  timed updates after initialization; $M_{\max}=32$).}
  \label{tab:c_throughput}
  \small
  \setlength{\tabcolsep}{5pt}
  \begin{tabular}{rrrrrrrr}
    \toprule
    \textbf{Window $n$} & \textbf{Pairs $N$} & \textbf{Invariant
    $\rho_n$} & \textbf{Updates/s} & \textbf{Latency} & \textbf{Zero}
    & \textbf{Heap} & \textbf{Fallback} \\
    \midrule
    25  &    300 & 92.0\% & \textbf{2,389,174} & 0.42 $\mu$s & 11.5\%
    & 88.5\% & 0.0\% \\
    35  &    595 & 94.3\% & \textbf{1,769,449} & 0.57 $\mu$s &  9.4\%
    & 90.6\% & 0.0\% \\
    51  &  1,275 & 96.1\% & \textbf{1,233,436} & 0.81 $\mu$s &  7.4\%
    & 92.6\% & 0.0\% \\
    75  &  2,775 & 97.3\% &   \textbf{875,130} & 1.14 $\mu$s &  5.7\%
    & 94.3\% & 0.0\% \\
    101 &  5,050 & 98.0\% &   \textbf{591,041} & 1.69 $\mu$s &  4.4\%
    & 93.4\% & 2.1\% \\
    151 & 11,325 & 98.7\% &   \textbf{244,784} & 4.09 $\mu$s &  3.3\%
    & 81.7\% & 15.0\% \\
    201 & 20,100 & 99.0\% &   \textbf{138,879} & 7.20 $\mu$s &  2.5\%
    & 71.3\% & 26.2\% \\
    \bottomrule
  \end{tabular}
\end{table}

Table~\ref{tab:c_throughput} shows that for $n \le 75$, $100\%$ of
updates are resolved by Tier~1 or Tier~2 in under $1.15\,\mu\text{s}$
per update. At $n=151$ and $n=201$, where the rank shift $m$ exceeds
$M_{\max}=32$ on $15.0\%$ and $26.2\%$ of steps, the hybrid Tier~3
frontier-guided matrix selector keeps average latency to
$4.09\,\mu\text{s}$ and $7.20\,\mu\text{s}$, respectively (a
  $35\times$ speedup at $n=201$ compared with a naive
$\mathcal{O}(n^2\log n)$ fallback sort).

\subsection{Repeated Implementation-Level Head-to-Head Benchmark}

We compared our exact implementation against the original native
\textsf{FQN} and \textsf{AFQN} implementations on the same machine.
For each window size, all three programs received the same saved
normal stream, consisting of $n$ initialization observations followed
by $100,000$ timed online updates (one warm-up and five measured
repetitions per program and window size, seed $20261113$).
Table~\ref{tab:head_to_head} reports median throughput from each
program's internal timer (\texttt{CLOCK\_MONOTONIC} for our code;
\texttt{gettimeofday} for \textsf{FQN} and \textsf{AFQN}) alongside
end-to-end process wall-clock throughput from a common external
\texttt{perf\_counter} timer.

\begin{table}[htbp]
  \centering
  \caption{Median throughput in the repeated implementation-level
    benchmark, rounded to three significant digits. Each entry is
    updates/s; internal and end-to-end wall-clock measurements are
  reported separately.}
  \label{tab:head_to_head}
  \small
  \setlength{\tabcolsep}{3pt}
  \begin{tabular}{r|rrr|rrr}
    \toprule
    & \multicolumn{3}{c|}{\textbf{Internal timer}} &
    \multicolumn{3}{c}{\textbf{End-to-end wall clock}} \\
    \textbf{Window $n$} & \textbf{Local exact} & \textbf{FQN} &
    \textbf{AFQN} & \textbf{Local exact} & \textbf{FQN} & \textbf{AFQN} \\
    \midrule
    25  & \textbf{2,220,000} & 490,000 & 1,330,000 &
    \textbf{1,870,000} & 439,000 & 693,000 \\
    35  & \textbf{1,670,000} & 313,000 &   779,000 &
    \textbf{1,450,000} & 291,000 & 496,000 \\
    51  & \textbf{1,140,000} & 205,000 &   550,000 &
    \textbf{1,020,000} & 195,000 & 393,000 \\
    75  &   \textbf{821,000} & 129,000 &   305,000 &
    \textbf{763,000} & 125,000 & 252,000 \\
    101 &   \textbf{552,000} & 103,000 &   228,000 &
    \textbf{525,000} & 100,000 & 196,000 \\
    151 &   \textbf{228,000} &  60,300 &   127,000 &
    \textbf{221,000} &  59,500 & 116,000 \\
    201 &   \textbf{127,000} &  51,400 &    99,400 &
    \textbf{125,000} &  50,800 &  92,800 \\
    \bottomrule
  \end{tabular}
\end{table}

Across all tested window sizes $n \in [25, 201]$, our exact
implementation outperforms both the exact \textsf{FQN} baseline (by
  $2.48\times$ to $6.35\times$ in internal throughput and $2.47\times$
to $6.09\times$ in wall-clock throughput) and the approximate
\textsf{AFQN} sketch (by $1.28\times$ to $2.69\times$ internally and
$1.35\times$ to $3.02\times$ wall-clock). The interquartile ranges of
the five internal-throughput measurements at $n=51$ were
$1{,}120{,}000$--$1{,}157{,}000$, $204{,}300$--$205{,}200$, and
$549{,}500$--$551{,}700$ updates/s for our implementation,
\textsf{FQN}, and \textsf{AFQN}, respectively; at $n=201$, they were
$126{,}300$--$128{,}300$, $51{,}100$--$51{,}600$, and
$99{,}100$--$99{,}900$ updates/s. In the head-to-head stream, the
recorded Tier~1 / Tier~2 / Tier~3 percentages were $7.4\% / 92.6\% /
0.0\%$ at $n=51$, $4.5\% / 93.4\% / 2.1\%$ at $n=101$, and $2.7\% /
71.6\% / 25.7\%$ at $n=201$.

To ensure functional parity with \textsf{FQN} and \textsf{AFQN}, our
head-to-head runner also reads the rolling median from
\texttt{sorted\_x} in $\mathcal{O}(1)$ time and evaluates the
$3\,Q_n$ outlier threshold on every timed step; \textsf{AFQN}
differs in that it computes a sketch approximation rather than the
exact order statistic. Because we benchmark the unmodified upstream
reference codes for reproducibility, implementation-level overhead
in those baselines (such as per-recursion heap allocations and
  duplicate $k_1=k_2$ selection in \textsf{FQN}'s \texttt{biselectV},
  or the tree-based \texttt{std::map} sketch container in
\textsf{AFQN}) contributes to their measured runtime; optimizing
these data structures could improve their absolute throughput,
though without eliminating the asymptotic per-step matrix biselection
in \textsf{FQN} or the $\mathcal{O}(n)$ logarithm and bucket updates
in \textsf{AFQN}.

To verify that this throughput advantage is not an artifact of
stationary Gaussian data, we repeated the same 5-repetition protocol
($T = 100,000$ updates, seed $20261113$) for $n \in \{51, 101, 201\}$
on three non-stationary or contaminated stress streams: an abrupt
$+10\sigma$ level shift at the midpoint, a contiguous $30$-observation
$+50\sigma$ outlier burst, and periodic $+25\sigma$ contamination
every $100$th observation (Table~\ref{tab:head_to_head_stress}). Across
all three regimes and all tested window sizes, our exact
implementation remains faster than both \textsf{FQN} ($2.31\times$ to
$5.84\times$ internally) and \textsf{AFQN} ($1.15\times$ to
$2.40\times$ internally).

\begin{table}[htbp]
  \centering
  \caption{Median internal and wall-clock throughput (updates/s,
    rounded to three significant digits) on non-stationary and
  contaminated $100,000$-update streams.}
  \label{tab:head_to_head_stress}
  \small
  \setlength{\tabcolsep}{3.5pt}
  \begin{tabular}{l r | rrr | rrr}
    \toprule
    & & \multicolumn{3}{c|}{\textbf{Internal timer}} &
    \multicolumn{3}{c}{\textbf{End-to-end wall clock}} \\
    \textbf{Scenario} & \textbf{$n$} & \textbf{Local exact} &
    \textbf{FQN} & \textbf{AFQN} & \textbf{Local exact} &
    \textbf{FQN} & \textbf{AFQN} \\
    \midrule
    Level shift ($+10\sigma$)     &  51 & \textbf{1,130,000} &
    201,000 & 563,000 & \textbf{1,020,000} & 192,000 & 404,000 \\
    & 101 &   \textbf{545,000} & 101,000 & 230,000 &
    \textbf{519,000} &  99,000 & 198,000 \\
    & 201 &   \textbf{125,000} &  50,600 &  99,800 &
    \textbf{123,000} &  50,000 &  93,100 \\
    \midrule
    Outlier burst ($+50\sigma$)   &  51 & \textbf{1,140,000} &
    203,000 & 556,000 & \textbf{1,030,000} & 194,000 & 399,000 \\
    & 101 &   \textbf{554,000} & 100,000 & 231,000 &
    \textbf{526,000} &  97,900 & 199,000 \\
    & 201 &   \textbf{126,000} &  51,300 & 100,000 &
    \textbf{125,000} &  50,700 &  93,600 \\
    \midrule
    Contaminated ($+25\sigma$)    &  51 & \textbf{1,160,000} &
    198,000 & 559,000 & \textbf{1,050,000} & 189,000 & 406,000 \\
    & 101 &   \textbf{528,000} & 100,000 & 227,000 &
    \textbf{502,000} &  98,100 & 195,000 \\
    & 201 &   \textbf{112,000} &  48,800 &  97,700 &
    \textbf{111,000} &  48,200 &  91,000 \\
    \bottomrule
  \end{tabular}
\end{table}

\begin{figure}[tbp]
  \centering
  \begin{tikzpicture}
    \begin{axis}[
        name=ax1,
        width=0.47\textwidth,
        height=5.2cm,
        ymode=log,
        xlabel={Window size $n$},
        ylabel={Median throughput (updates/s)},
        xtick={25,51,75,101,151,201},
        xmin=18, xmax=210,
        ymin=40000, ymax=3500000,
        grid=both,
        major grid style={gray!25},
        minor grid style={gray!10},
        tick label style={font=\scriptsize},
        label style={font=\footnotesize},
        title style={font=\footnotesize\bfseries},
        title={(a) Head-to-head throughput (log scale)},
        legend style={at={(0.97,0.97)}, anchor=north east,
        font=\scriptsize, fill=white, fill opacity=0.85, draw=gray!50},
      ]
      \addplot[thick, blue!75!black, mark=*, mark size=1.8pt] coordinates {
        (25,2221482) (35,1672269) (51,1140017) (75,820998)
        (101,552242) (151,227761) (201,127285)
      };
      \addlegendentry{Local exact}

      \addplot[thick, dashed, orange!85!black, mark=square*, mark
      size=1.7pt] coordinates {
        (25,1325926) (35,779302) (51,549731) (75,305423) (101,227611)
        (151,126920) (201,99424)
      };
      \addlegendentry{\textsf{AFQN}}

      \addplot[thick, dotted, red!75!black, mark=triangle*, mark
      size=2.0pt] coordinates {
        (25,489860) (35,313293) (51,204742) (75,129196) (101,102752)
        (151,60338) (201,51389)
      };
      \addlegendentry{\textsf{FQN}}
    \end{axis}

    \begin{axis}[
        at={(ax1.south east)},
        anchor=south west,
        xshift=1.35cm,
        width=0.46\textwidth,
        height=5.2cm,
        ybar stacked,
        bar width=9pt,
        symbolic x coords={25,35,51,75,101,151,201},
        xtick=data,
        xlabel={Window size $n$},
        ylabel={Share of updates (\%)},
        ymin=0, ymax=100,
        ymajorgrids=true,
        grid style={gray!25},
        tick label style={font=\scriptsize},
        label style={font=\footnotesize},
        title style={font=\footnotesize\bfseries},
        title={(b) Execution-tier profile (Table~\ref{tab:c_throughput})},
        legend style={at={(0.03,0.97)}, anchor=north west,
        font=\scriptsize, fill=white, fill opacity=0.9, draw=gray!50},
      ]
      \addplot[fill=green!55!black!70, draw=green!45!black] coordinates {
        (25,11.5) (35,9.4) (51,7.4) (75,5.7) (101,4.4) (151,3.3) (201,2.5)
      };
      \addlegendentry{Tier 1 (Zero-step)}

      \addplot[fill=blue!55!white, draw=blue!65!black] coordinates {
        (25,88.5) (35,90.6) (51,92.6) (75,94.3) (101,93.4) (151,81.7) (201,71.3)
      };
      \addlegendentry{Tier 2 (Row-heap)}

      \addplot[fill=orange!75!white, draw=orange!75!black] coordinates {
        (25,0.0) (35,0.0) (51,0.0) (75,0.0) (101,2.1) (151,15.0) (201,26.2)
      };
      \addlegendentry{Tier 3 (Frontier)}
    \end{axis}
  \end{tikzpicture}
  \caption{Empirical scaling and tier profile on the $100,000$-update
    Gaussian stream. (a)~Median internal throughput from
    Table~\ref{tab:head_to_head} on a logarithmic scale. (b)~Percentage
    of updates resolved in each tier from Table~\ref{tab:c_throughput}
  ($M_{\max} = 32$).}
  \label{fig:benchmark_scaling}
\end{figure}
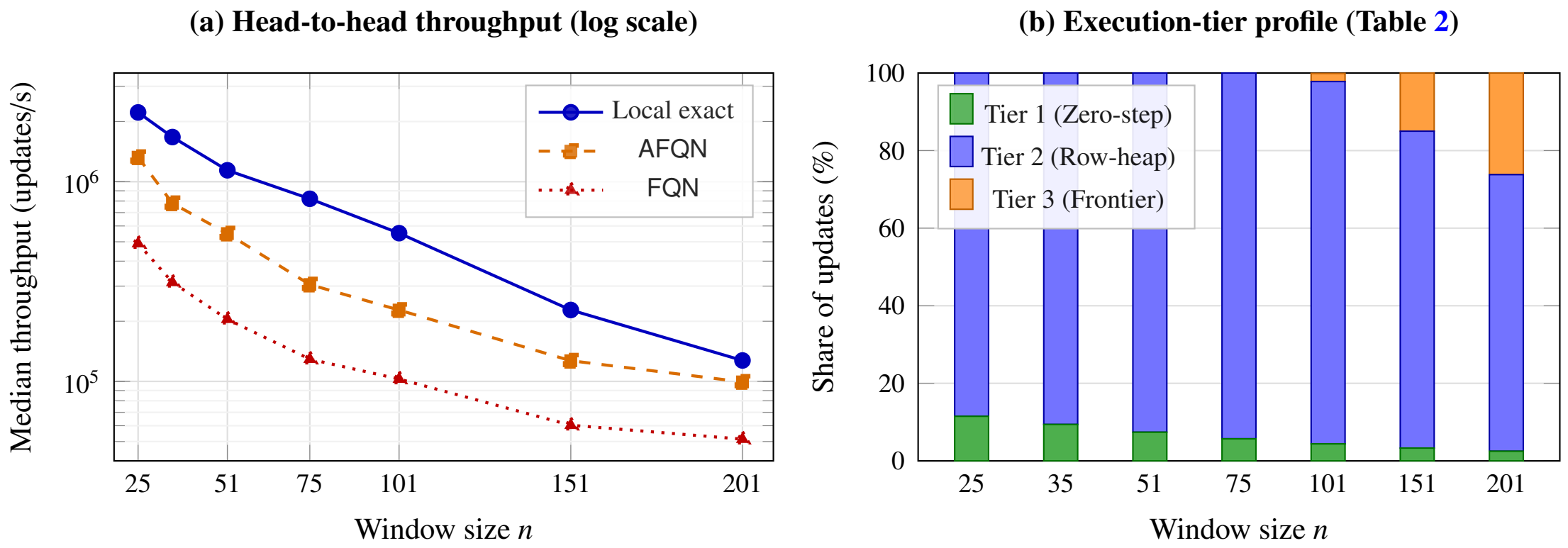

Figure~\ref{fig:benchmark_scaling} summarizes the Gaussian
head-to-head throughput scaling alongside the execution-tier profile
from Table~\ref{tab:c_throughput}, illustrating how Tier~3 activation
above $n=101$ shapes the large-window slope while preserving a clear
advantage over both \textsf{FQN} and \textsf{AFQN}.

\section{Software Availability}
\label{sec:software}

The self-contained ANSI C reference implementation for exact online
$Q_n$ estimation, together with the batch routine and
benchmark-reproduction scripts used in this study, is available from
the author upon reasonable request. Companion R and Python package
prototypes are in preparation.

\section{Conclusions}
\label{sec:conclusion}

We have presented a hybrid exact rank-update framework for
sliding-window $Q_n$ estimation and evaluated an ANSI C
implementation on synthetic streams. Our theoretical analysis
establishes tie-exact $\mathcal{O}(\log n)$ dual-rank verification,
$\mathcal{O}(n + m\log n)$ localized row-heap extraction via
monotonic two-pointer frontier scanning, and $\mathcal{O}(n\log n)$
frontier-guided implicit matrix selection for larger rank shifts ($m
> M_{\max}$), without materializing the $\binom{n}{2}$ pairwise
difference matrix. Multi-regime stress tests and exhaustive boundary
checks confirm zero double-precision discrepancy against an
independent offline batch reference. In repeated same-platform
benchmarks on a Gaussian stream across $n \in [25, 201]$, our exact
implementation consistently outperforms both the exact \textsf{FQN}
baseline (by $2.5\times$ to $6.4\times$) and the approximate
\textsf{AFQN} sketch (by $1.3\times$ to $2.7\times$). Future work
includes extending the dual-rank invariance paradigm to online
generalized median estimators for subset sizes $k \ge 3$, where
sliding-window overlap drops to $1 - k/n$ and lacks two-dimensional
matrix monotonicity.

\appendix

\section{Proofs of Theoretical Results}
\label{app:proofs}

\subsection*{Proof of Lemma~\ref{lem:invariance}}
\begin{proof}
  Every pair of distinct observation occurrences in $W_t$ either
  includes $x_{\text{old}}$ ($n-1$ pairs, forming $D_{\text{evict}}$)
  or is drawn exclusively from the $n-1$ surviving occurrences
  ($\binom{n-1}{2}$ pairs). Dividing $\binom{n-1}{2} =
  \frac{(n-1)(n-2)}{2}$ by $\binom{n}{2} = \frac{n(n-1)}{2}$ yields
  the invariance ratio $\rho_n = \frac{n-2}{n} = 1 - \frac{2}{n}$.
  Because transitioning from $D(W_t)$ to $D(W_{t+1})$ removes at most
  $n-1$ differences smaller than $\theta$ and inserts at most $n-1$
  such differences, the net change in both $R_{<}(\theta; \cdot)$ and
  $R_{\le}(\theta; \cdot)$ is bounded within $[-(n-1), \, +(n-1)]$.
\end{proof}

\subsection*{Proof of Theorem~\ref{thm:dual_rank}}
\begin{proof}
  By Lemma~\ref{lem:invariance}, $D(W_t) = D(W_{\text{shared}})
  \uplus D_{\text{evict}}$ and $D(W_{t+1}) = D(W_{\text{shared}})
  \uplus D_{\text{ingest}}$. By additivity of multiset cardinality:
  \begin{align*}
    R_{<}(\theta; W_{t+1}) &= R_{<}(\theta; W_t) -
    C_{<}(D_{\text{evict}}, \theta) + C_{<}(D_{\text{ingest}}, \theta), \\
    R_{\le}(\theta; W_{t+1}) &= R_{\le}(\theta; W_t) -
    C_{\le}(D_{\text{evict}}, \theta) + C_{\le}(D_{\text{ingest}}, \theta).
  \end{align*}
  These identities yield Eqs.~\eqref{eq:update_lt}--\eqref{eq:update_le}.

  To prove Eqs.~\eqref{eq:bsearch_lt}--\eqref{eq:bsearch_le}, observe
  that for any $j \in \{1, \dots, n\}$, $|x_{(j)} - v| < \theta$ is
  equivalent to $v - \theta < x_{(j)} < v + \theta$. Because $x_{(1)}
  \le \dots \le x_{(n)}$ is monotonically non-decreasing, the set of
  indices satisfying $x_{(j)} > v - \theta$ is $\{U_{>}(v - \theta),
  \dots, n\}$, and the set of indices satisfying $x_{(j)} < v +
  \theta$ is $\{1, \dots, U_{\ge}(v + \theta) - 1\}$. Their
  intersection has cardinality $U_{\ge}(v + \theta) - U_{>}(v -
  \theta)$ whenever $\theta > 0$. Since $v = x_{(p)}$ itself
  satisfies $|x_{(p)} - v| = 0 < \theta$ whenever $\theta > 0$,
  subtracting $\mathbf{1}_{\{\theta > 0\}}$ excludes the self-pair
  $(p, p)$, establishing Eq.~\eqref{eq:bsearch_lt}. Similarly,
  $|x_{(j)} - v| \le \theta \iff v - \theta \le x_{(j)} \le v +
  \theta$ corresponds to the index interval $[U_{\ge}(v - \theta), \,
  U_{>}(v + \theta) - 1]$, which always contains $p$ since $\theta
  \ge 0$; subtracting $1$ excludes the self-pair and proves
  Eq.~\eqref{eq:bsearch_le}. Each bound $U_{\ge}, U_{>}$ is resolved
  via binary search in $\lceil \log_2(n+1) \rceil$ comparisons.
\end{proof}

\subsection*{Proof of Theorem~\ref{thm:heap_correctness}}
\begin{proof}
  Consider Case~1 ($R_{\le}(\theta; W_{t+1}) < k$). By definition of
  $R_{\le}$, exactly $R_{\le}$ occurrences of the multiset
  $D(W_{t+1})$ are $\le \theta$, and the remaining $N - R_{\le} \ge k
  - R_{\le} = m$ occurrences form the multiset
  \[
    D_{>}(\theta) = \{\!\{ x_{(j)} - x_{(i)} : 1 \le i < j \le
    n,\ x_{(j)} - x_{(i)} > \theta \}\!\}.
  \]
  Therefore, the $k$-th smallest occurrence in $D(W_{t+1})$ is
  precisely the $m$-th smallest occurrence in $D_{>}(\theta)$.

  In each row $i \in \{1, \dots, n-1\}$, the sequence of differences
  $d_{i,j} = x_{(j)} - x_{(i)}$ for $j = i+1, \dots, n$ is
  monotonically non-decreasing ($d_{i, i+1} \le d_{i, i+2} \le \dots
  \le d_{i, n}$). Hence, the occurrences in row $i$ belonging to
  $D_{>}(\theta)$ form the contiguous suffix with entries $d_{i,
  j^+(i)}, d_{i, j^+(i)+1}, \dots, d_{i, n}$, with minimum $d_{i,
  j^+(i)}$. Moreover, because $x_{(i)} \le x_{(i+1)}$, we have
  $x_{(j)} - x_{(i+1)} \le x_{(j)} - x_{(i)}$ for every column $j$,
  which implies $j^+(i+1) \ge \max(i+2, j^+(i))$. Thus, all non-empty
  frontier indices $j^+(1), \dots, j^+(n-1)$ are located in
  $\mathcal{O}(n)$ comparisons via a single monotonic two-pointer scan.

  Heap entries include the row and column indices, so equal
  difference values from distinct pairs remain separate occurrences.

  We prove by induction on step $s \in \{1, \dots, m\}$ that the
  $s$-th value popped from $\mathcal{H}_{\min}$ is the $s$-th
  smallest element of $D_{>}(\theta)$. At $s=1$, $\mathcal{H}_{\min}$
  contains the minimum element of every non-empty row suffix of
  $D_{>}(\theta)$; since the global minimum of a union of sorted
  sequences must be the minimum of at least one sequence, the first
  popped value is the smallest element of $D_{>}(\theta)$.
  Inductively, after $s-1$ pops, removing the head of a row suffix
  and inserting its immediate successor $d_{i, j+1}$ maintains the
  invariant that $\mathcal{H}_{\min}$ holds the minimum unextracted
  element of every active row suffix in $D_{>}(\theta)$. Thus, the
  $m$-th popped value is the $m$-th smallest element of
  $D_{>}(\theta)$, which equals $D(W_{t+1})_{(R_{\le} + m)} = D(W_{t+1})_{(k)}$.

  Case~2 ($R_{<}(\theta; W_{t+1}) \ge k$) follows by order duality:
  since $R_{<}$ occurrences of $D(W_{t+1})$ are strictly less than
  $\theta$, the $k$-th smallest occurrence of $D(W_{t+1})$ is the
  $(R_{<} - k + 1)$-th largest occurrence of the multiset
  \[
    D_{<}(\theta) = \{\!\{ x_{(j)} - x_{(i)} : 1 \le i < j \le
    n,\ x_{(j)} - x_{(i)} < \theta \}\!\},
  \]
  and the frontier $j^-(i) = \min\{j > i : x_{(j)} - x_{(i)} \ge
  \theta\} - 1$ is likewise non-decreasing with $i$. Monotonic
  frontier scanning and Floyd bottom-up heap construction each take
  $\mathcal{O}(n)$ comparisons, and $m$ extractions take
  $\mathcal{O}(m \log n)$ comparisons, totaling $\mathcal{O}(n + m \log n)$.
\end{proof}

\bibliographystyle{elsarticle-harv}
\bibliography{ExactOnlineQn}

\end{document}